\documentclass{epac98}
\usepackage{epsfig}

\newcommand{\ff}{\mbox{${\cal F} 1$}}

\newcommand{\mum}{$\mu$m}

\title{TDC Chip and Readout Driver Developments \\ for COMPASS
       and LHC-Experiments}

\author{
 G.~Braun,
\underline{H.~Fischer}$^\ast$,
J.~Franz, A.~Gr\"unemaier, F.H.~Heinsius, K.~K\"onigsmann, \\[-0.05cm]
M.~Schierloh, T.~Schmidt, H.~Schmitt, H.J.~Urban \\[0.1cm]
 Universit\"at Freiburg, \\[-0.05cm]
Fakult\"at f\"ur Physik, Hermann-Herder-Str. 3,
D-79104 Freiburg, Germany\\
{\tt \normalsize $^\ast$EMAIL: Horst.Fischer@cern.ch} 
\\[0.1cm]
Contribution to: \\[-0.05cm] {\it fourth workshop on electronics for 
LHC experiments - LEB98} \\[-0.05cm] Rome, 21-25 September 1998. 
}

\begin{document}

\maketitle

\abstract{
A new TDC-chip is under development for the COMPASS experiment at CERN.
The ASIC, which exploits the $0.6  \mu \mathrm{m}$\  CMOS sea-of-gate
technology, will allow high resolution time measurements
with digitization of
$75 \; \mathrm{ps}$, and an unprecedented degree of flexibility
accompanied by high rate capability and low power consumption.
Preliminary specifications of this new TDC chip are presented.

Furthermore a FPGA based readout-driver and buffer-module as an
interface between the front-end of the COMPASS detector systems
and an optical S-LINK is in development. The same module serves also
as remote
fan-out  for the COMPASS trigger distribution and time synchronization
system. This readout-driver monitors the trigger and data flow to and
from front-ends. In addition, a specific data buffer structure
and sophisticated data flow control is used  to pursue local pre-event
building. At start-up the module controls all necessary front-end
initializations. }

\section{Introduction}

The COMPASS experiment at CERN will investigate  hadron structure
by deep inelastic scattering processes and in addition pursue
different aspects of hadron spectroscopy using hadron beams.
Comparison with calculations based on operator product expansion
or lattice techniques and with model predictions  based on chiral
symmetry or effective degrees of freedom will help to improve our
understanding of hadrons.  To reach this objective a new
state-of-the-art fixed target spectrometer, capable of standing
beam intensities of up to $2\cdot 10^8$\ particles/spill and with
excellent particle identification, will be put into commission in
the year 2000.

These challenging physics goals of the COMPASS Experiment can only be met,
if at highest possible beam rates large data statistics can be recorded.
This leads to the requirement of an experiment with
negligible dead time, which can digest data rates of several Gigabyte
per second. This is at the edge of today's digitization and bandwidth
technologies.

The initial approach of the experiment was to minimize dead time by
reading digitized data through pipelines where applicable. Availability
of and resolution requirements on digitization units
have made this goal not feasible
for every detector component. The present scenario assumes a maximum
dead time of 500~ns for analog sensitive front-end electronics which
will digitize signal amplitude or charge with a
precision better than 10~bit.

\begin{figure}[htb]
\centering
\epsfig{file=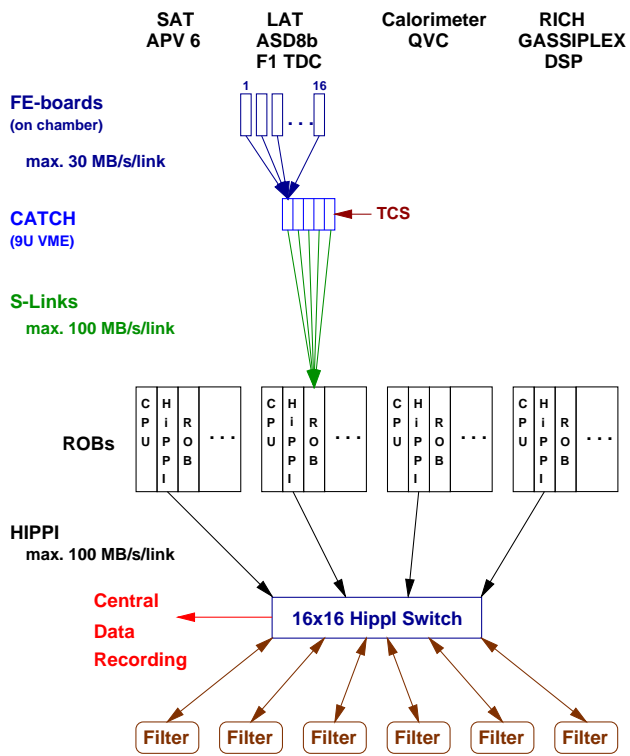, width=82.5mm}
\caption{The architecture of the COMPASS detector readout.}
\label{fig:architecture}
\end{figure}

The readout architecture of the COMPASS experiment is summarized in
Figure~\ref{fig:architecture}. Data are digitized right at the detector
by the front-end electronics wherever possible. In case of
analog readout the pedestal subtraction and zero suppression is
performed by the front-end at the detector.
To suppress background for time measurements only those hits
are transfered to the data recording units which have a
correlation to a trigger time.

The data are  transmitted from the front-end to
CATCH modules (CATCH = COMPASS
Accumulate, Transfer and Cache Hardware). The functionality of the
CATCH will be described below.

From the CATCH data are transmitted via a standardized link to the
read-out buffers (ROB) which can store all data from one spill.
The backbone of this data transfer is the S-link~\cite{slink}
interface and the S-LINK data transfer protocol.
Presently about 128 S-LINK connections are
foreseen to transmit data with a maximum total bandwidth of
about 12~GB/s.

The read-out buffers combine data which belong to one event,
check consistency of the data and perform sub-event-building.
In a next step they transmit the sub-events via
HiPPI~\cite{hippi} or Gigabit Ethernet to filter computers. Here the
final event-building is performed and events are reconstructed.
The filter farm will reduce the data based on physics cuts by
a factor of 5
to 10 and a continuous rate of 12 to 24 MB/s will be transfered
to the central data recording facilities at CERN.

\section{The \ff\ TDC Chip}

A key element in the experiment is a new developed dead time free
TDC chip, the \ff. It will be used to digitize the data from a
large majority of detectors. Table~\ref{tab:appllist} shows that
the required specifications vary significantly for the foreseen
applications of the \ff\ chip. Certainly the most challenging
demands are defined by the scintillating fiber (SCIFI) detectors.
Count rates up to 6 MHz per channel and time resolution better
than 100 ps are on the edge of today's technology. When count
rates and background become too high, existing  limitations in
bandwidth on data-links between TDC and the data recording system
require on-chip event selection. Detector components like Plastic
Iarocci Tubes ($\mu$-Detector~1) or MWPC do not ask for precise
timing but require lowest cost/channel  due to the large number of
channels involved. Thus multiplexed input is desirable for this
kind of detectors. We were forced to develop a new multi-purpose
TDC Chip, because no integrated TDC circuit is currently on the
market with a sufficient degree of integration and flexibility to
fulfill all requirements of COMPASS. In spite of the short time
available for the R\&D phase for COMPASS we decided to develop a
TDC in collaboration with {\it acam}~\cite{acam}, a recently
founded German enterprise specialized in the development of time
measurement devices and applications.

\begin{table}[htb]
\caption{ Possible applications for the \ff\ TDC-chip
in the COMPASS experiment.}
\label{tab:appllist}\medskip
\begin{center}
{\small
\begin{tabular}{|c|c|c|c|c|c|}
\hline
detector &  \hspace*{-0.2cm} channels \hspace*{-0.2cm}  & time
         & \hspace*{-0.15cm}  background\hspace*{-0.15cm}   & event\\
                    &
         & \hspace*{-0.2cm} resolution \hspace*{-0.2cm}   & rate
         & rate \\
                    &           & $[$ps$]$   & $[$kHz$]$&$[$kHz$]$\\
\hline
Straws              & 37000     &   1000     &    600    &  100 \\
\hline
$ \hspace*{-0.15cm}  \mu$Mega Chambers \hspace*{-0.15cm}
                    & 15000     & $<$500     &  $<$100   & $<$100 \\
\hline
$\mu$-Detector 1 \hspace*{-0.2cm}
                    &  9000     &   ---      &     50    & $<$1 \\
\hline
$\mu$-Detector 2 \hspace*{-0.2cm}
                    &  3000     &   2000     &    130    & $<$1 \\
\hline
  MWPC              & 18000     &   ---      &    600    & 100  \\
\hline
 SCIFI              &  1280     &    320     &    600    & 6000 \\
\hline
  \hspace*{-0.15cm}Beam Scintillators    \hspace*{-0.15cm}
                     &   512     &  $<$100    &    600    & 6000 \\
\hline
 \hspace*{-0.2cm} Recoil-Detector  \hspace*{-0.2cm}
                    & 5000      &  $<$100    & $<$100    & $<$100 \\
\hline
Hodoscope           &  1200     &  $<$100    & $<$100    & $<$100 \\
\hline

\end{tabular}
}
\end{center}
\end{table}

The development of the \ff\ resulted in a highly programmable
eight channel TDC ASIC based on a low cost $0.6$\mum\ CMOS
sea-of-gates process. The reasons for only eight channels in a
single chip are driven by financial considerations as well as
physical limitations. Eight channels in a chip optimize savings
due to density on the die and costs for bonding and chip package.
In addition chip prizes depend strongly  on the number of produced
chips and less on the die surface. Hence 6000 chips with  8
channels are more cost efficient than 1500 chips with 32 channels.

The scheme utilized in the \ff\  avoids the use of high-speed clocks
and shift registers and, therefore, results in very low power
consumption. Signal propagation times in gate arrays
have very large variations as a function  of temperature and operation
voltage. Furthermore signal propagation within gates have a spread
strongly dependent on the production process.
To overcome these disadvantages the \ff\ exploits a self calibrating
method utilized by voltage controlled delay elements as part of an
asymmetric oscillator in a phase locked loop (PLL).

\begin{figure}[htb]
\centering
\epsfig{file=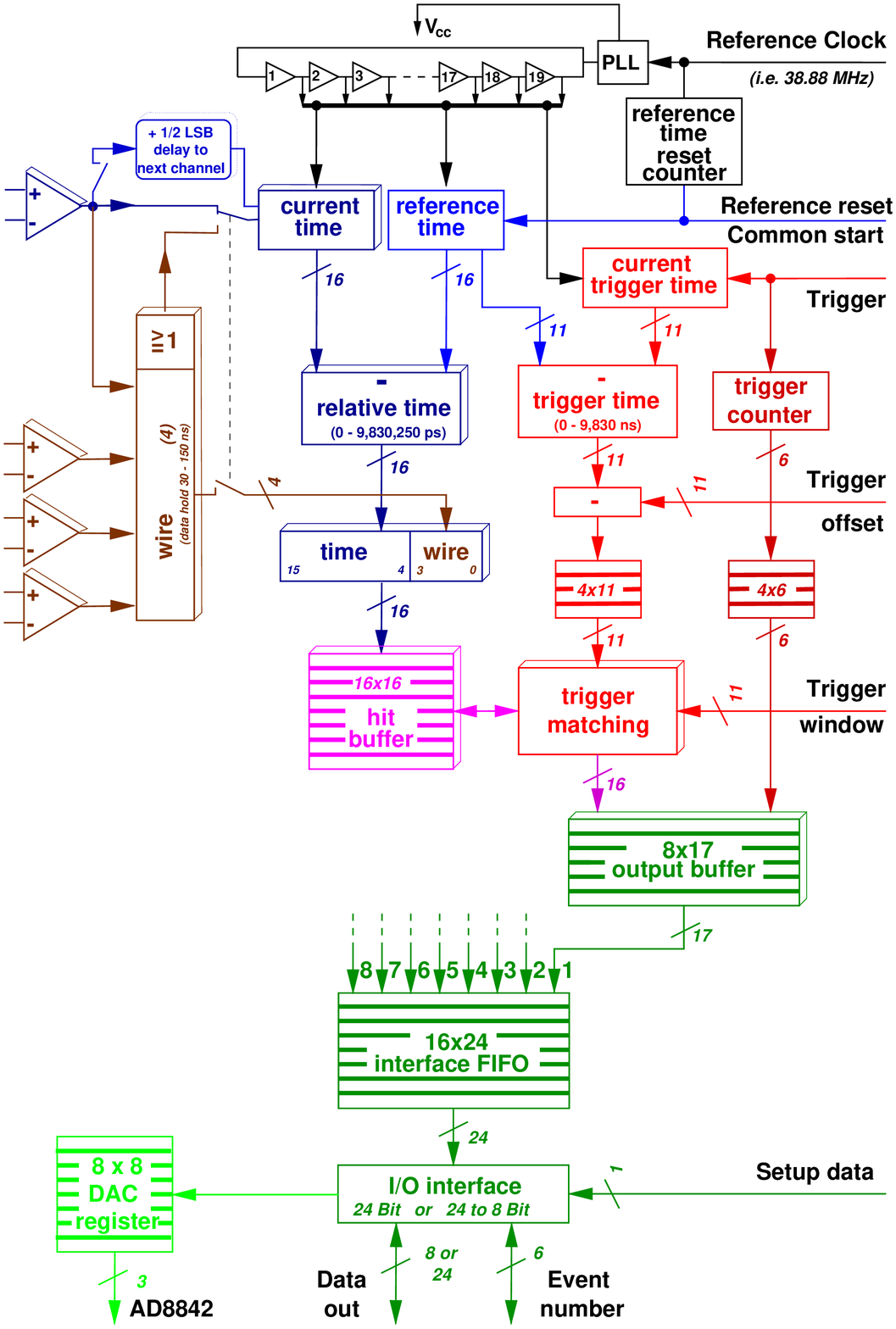, width=82.5mm}
\caption{Outline of the \ff\ chip.}
\label{f1dia}
\end{figure}

An outline of the chip is shown in Fig.~\ref{f1dia}. Non-perspective
displayed boxes - like trigger counter, PLL or interface FIFO - refer
to units which are shared between the eight channels, whereas the
perspective displayed boxes - like hit buffer, trigger matching and
readout buffer - indicate logic units in the chip which come eightfold
and are not shared between the channels.

The preliminary specifications for the TDC are listed in
Table~\ref{techlist}. The \ff\ can be considered as a real multi-purpose
chip  with very different modes of operation in terms of resolution
and functionality.

\begin{table}
\caption{ Preliminary technical description of the \ff\ based on
pre-layout simulations. } \label{techlist} \medskip \begin{center}
\begin{tabular}{l} {\bf Number of channels:}  \\ \hspace*{0.2cm} 4
for high resolution mode \\ \hspace*{0.2cm} 8 for standard
resolution mode  \\ \hspace*{0.2cm}32 for pattern recognition mode
\\[3pt] {\bf Time bin size:}  \\ \hspace*{0.2cm}75 ps for high
resolution mode \\ \hspace*{0.2cm}150 ps for standard operation
mode \\ \hspace*{0.2cm}5700 ps for pattern recognition mode
\\[3pt] {\bf Reference-clock frequency:  }      \\
\hspace*{0.2cm}Between 500 kHz and 40 MHz  \\
\hspace*{0.2cm}(Clock is used for self calibration only) \\[3pt]
{\bf Differential non-linearity: } \\ \hspace*{0.2cm}Less than
0.05 LSB for high resolution mode\\ \hspace*{0.2cm}Less than 0.2
LSB for standard mode\\[3pt] {\bf Integral non-linearity: }   \\
\hspace*{0.2cm}Less than one time bin\\[3pt] {\bf Variation with
temperature: } \\ \hspace*{0.2cm}Less than one time bin \\[3pt]
{\bf Dynamic range:  }     \\ \hspace*{0.2cm}16 bits \\[3pt] {\bf
Double pulse resolution:}  \\ \hspace*{0.2cm}Typical 22 ns \\[3pt]
{\bf Digitization and readout dead time:}  \\ \hspace*{0.2cm}None
\\[3pt] {\bf Hit buffer size: }   \\ \hspace*{0.2cm}32
measurements for high resolution mode \\ \hspace*{0.2cm}16
measurements for standard mode \\ \hspace*{0.2cm}16 measurements
for pattern recognition mode\\[3pt] {\bf Output buffer size:}  \\
\hspace*{0.2cm}8 measurements  \\[3pt] {\bf Readout Interface: }
\\ \hspace*{0.2cm}16 measurements  \\[3pt] {\bf Trigger buffer
size: }   \\ \hspace*{0.2cm}4 \\[3pt] {\bf Power Supply: } \\
\hspace*{0.2cm}5.0 V, optional 3.3 V with reduced resolution\\
\hspace*{0.2cm}30 mA - 80 mA,  depending on trigger load \\[3pt]
{\bf Temperature range:  }  \\ \hspace*{0.2cm}-40 to  +85 degree
centigrade \\[3pt] {\bf Hit input: } \\ \hspace*{0.2cm}LVDS or TTL
\\[3pt] {\bf Package:} \\ \hspace*{0.2cm}160 PQFP \\[3pt]
\end{tabular}
\end{center}
\end{table}

In the standard mode the \ff\ is an eight channel TDC chip with a
digitization uncertainty of 150~ps. When a leading or trailing
edge of a LVDS-signal~\cite{lvds} is present at one of the channel
inputs the time relative to the last reset of the TDC is derived
from a PLL and a coarse counter. The time stamp is written into
the hit buffer. When a TTL signal edge is present at the trigger
input the time relative to the last reset is measured for the
trigger signal. In a next step the time stamps, which are stored
in the hit buffer, are compared to the time stamp of the trigger
time. Data within a pre-set time window are accepted for readout
and copied to the output buffer.  An 8 or alternatively 24 bit
parallel bus is used for readout of the chip. The data bus can be
operated at a maximum frequency of 50 MHz.

In the high resolution mode two channels are interleaved by 1/2
least significant bit (LSB)
delay. Hence in this mode the \ff\ has  four channels with a resolution
of 75~ps and the hit buffer for a single channel
is increased by a factor of two. Otherwise the principle of operation
is the same as in the standard mode.

The third way to operate the chip is the pattern recognition or
latch mode. In this
mode 32 input channels are connected to eight groups of fourfold
latch-registers. Once a leading edge of a signal is detected on a
input channel the fourfold input register is activated for the duration
of a pre-set time interval and subsequent hits on the other inputs can be
registered as well. When the pre-set time has passed, the register is
copied to the hit buffer and cleared for new recording. While this
transfer is realized a time-stamp with reduced resolution is taken and
added to the information from the register.

In the fourth mode the chip can be used like any other normal
common-start TDC. After a start-signal on a dedicated start-input
has been applied all subsequent time measurements are relative to
this start-signal. In this mode the time-stamps are immediately
passed to the output bus and not stored in the hit buffer. This
mode does not make use of the trigger-matching possibilities of
the TDC.

Initialization of the \ff -chip is performed via a 10 Mbps serial
connection. The interface for this link is integrated in the TDC
chip and no additional external control is needed.

As an additional feature the \ff\ has eight eight-bit registers
and a dedicated interface to an eight channel digital-to-analog
converter (AD8842)~\cite{dac}, which can be used for threshold
control of discriminator units placed on the front-end cards. The
integration of the register and the interface in the TDC avoids
 additional memory  and controller units on the
front-end board.

\section{Readout Driver Boards for COMPASS}

The basic philosophy behind the COMPASS readout architecture
design was to unify as much hardware components as possible to
save on spare parts and manpower during preparation and during the
data-taking phase of the experiment. Having this concept in mind
we decided to concentrate the data as close as possible to the
front-end into few high-bandwidth data streams. A logic
consequence was the development of a standardized data multiplexer
board, the CATCH, for all the different components of the COMPASS
detector. The CATCH functions mainly as a derandomizer and must
provide enough memory for intermediate storage of several events.
To concentrate a large number of channels we decided to build the
CATCH as a 9U VME board. However, the VME bus will only be used
for power distribution, for transfer of set-up data during
detector initialization and to spy on a sub-sample of events
parasitically - but independent of the data acquisition system -
during data taking.

Data input to the CATCH can be performed in several ways. The
data-interfaces to the CATCH are designed as mezzanine cards
following the IEEE CMC standard~\cite{mezzanine}. The advantage of
mezzanine cards is that they allow easy and fast exchange without
hardware modifications and thus give the highest degree of
flexibility.  Currently we design four different types of
mezzanine cards: One card contains four 30 MByte/s HOTLink
de-serializer receiver-chips~\cite{HOTLink} which will be used for
front-end boards mounted on the detector; a second card contains
\ff -chips to assemble TDC boards; a third mezzanine card which
contains several FPGA to implement fast 200 MHz dead-time-free
scalers and a fourth card which will serve as an interface to the
silicon detector readout boards.

\begin{figure}[htb]
\centering
\epsfig{file=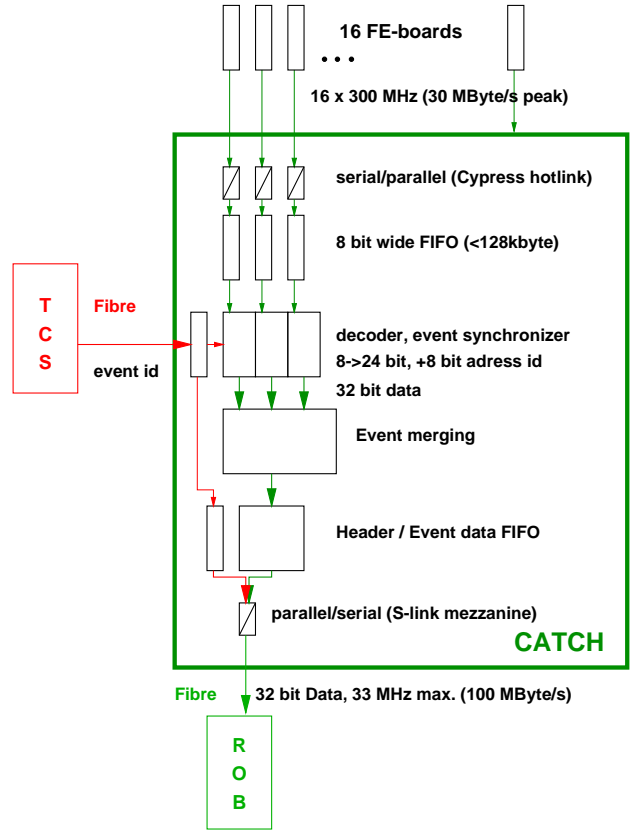, width=82.5mm}
\caption{Data flow in the CATCH module illustrated for
wire chamber readout via HOTLink receivers.}
\label{fig:catch-data}
\end{figure}

The output of the CATCH module is connected through the P2
connector to a S-LINK multiplexer board mounted on the backplane
of the VME crate. This multiplexer board is used to optimize the
transmission rate on the S-LINK to the available S-LINK bandwidth,
which is 100 MB/s/link. Figure~\ref{fig:catch-data} illustrates,
as an example for the data flow on the CATCH, the use of the
Cypress HOTLink receiver-chip as data input source. In this
example 24~bit data words are transmitted in three words of one
byte each.

The data interfaces are scanned continuously for arriving events.
All data are reformatted to 32-bit words and, if necessary,
additional hardware addresses are added. These are transfered by
the front-end only during the initialization process for
unambiguous identification. Next all data from a trigger are
packed between a S-LINK header and S-LINK trailer. The S-LINK
header and trailer contain a unique bit pattern for
identification. Furthermore a S-LINK header contains an event- and
spill-number and describes the data source and event type. For
each individual event the information contained in the header is
received by the CATCH from the Trigger Control System (TCS). The
data for one event are passed to the ROB when all front-end cards,
which are connected to a CATCH, have transmitted either valid-data
or an empty-data word for a particular trigger-number.

The CATCH module can also be used to erase data based on second
level decisions from the readout stream. To implement this feature
in an experiment the trigger numbers of events, which are rejected
by the higher trigger level, have to be distributed to the CATCH
boards via the trigger control system. The latency of the second
level trigger decision and the total data rate per CATCH
determines the memory needed to store events before they are
passed to the ROB. Currently the implementation of several data
compression methods on the level of the CATCH modules are
discussed as a possibility for further reduction of data rates.

\begin{figure}[bt] \centering
\epsfig{file=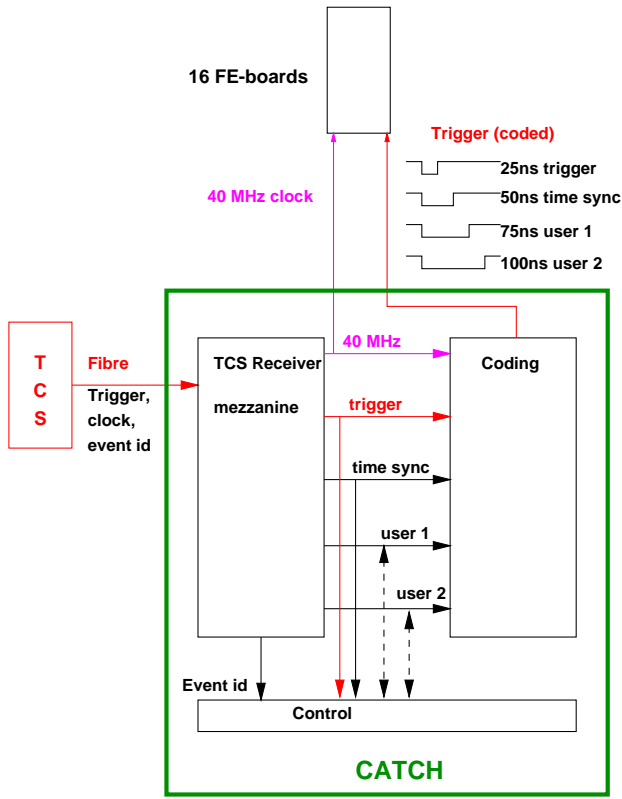, width=82.5mm} \caption{Trigger
flow in the CATCH module. } \label{fig:catch-trigger}
\end{figure}

The handling of trigger signals from TCS on the CATCH board is
illustrated in Figure~\ref{fig:catch-trigger}. The CATCH-module
receives trigger signals and a reference clock from the TCS via
optical data transmission. The TCS interface board contains an
optical or copper cable receiver and a unit to decode the
time-division multiplexed signals transmitted by the TCS. The TCS
receiver board can be mounted to the P2 connector on the backplane
of VME crates. Once the CATCH receives a trigger signal from the
TCS it encodes the signal as a pulse with the length of one clock
cycle. The encoded trigger signal is passed immediately to the
front-end cards synchronous to the COMPASS reference clock (38.88
MHz). To check data for consistency a local event number is
generated and compared to the event numbers transfered with the
data headers from the front-end.

The physical network layer between front-end and CATCH is realized
in standard S/STP~CAT~6 \cite{sutp} cable. This cable is widely
used in Ethernet links between wall outlets and personal
computers. The great demand for this kind of cable makes it the
cheapest solution to connect two points with four differential
lines up to frequencies of 700 MHz. In our application one
differential line is used to distribute the COMPASS 38.88~MHz
reference clock to the front-end. The second line is used to
distribute up to four different user signals to the front-end. The
signals are distinguished by their length with respect to the
COMPASS reference clock. One of the four possible signals is the
trigger-signal and another is a time synchronization reset at the
beginning of a spill. The remaining two can be defined according
to user requirements. A third line is used for as serial link to
down-load initialization data to the front-end electronics at
start-up. This serial line is operated at 10 Mbps. The fourth line
is used to transfer the serialized data from the HOTLink
transmitter (front-end) to the HOTLink receiver (CATCH).

\section{Acknowledgement}
This work would be impossible without the significant efforts by
the staffs of the collaborating institutions, in particular those
involved in front-end electronics development. The developments described
in this report are supported by the German Bundesministerium
f\"ur Bildung, Wissenschaft, Forschung und Technologie.

% \section*{References}

\end{document}